\documentclass{amsart}
\usepackage{amsmath}
\usepackage{graphicx}
\usepackage{dcolumn}
\usepackage{bm}
\usepackage{wrapfig}
\usepackage{epsfig}
\usepackage{breqn}

\theoremstyle{definition}

\theoremstyle{remark}

\numberwithin{equation}{section}

\begin{document}

\title[Vacuum Energy in a B-I Universe]{Exploring Vacuum Energy in a Two-Fluid Bianchi Type I Universe}
\title{Exploring Vacuum Energy in a Two-Fluid Bianchi Type I Universe}


\author{Ikjyot Singh Kohli}
\address{Department of Physics and Astronomy}
\curraddr{York University, Toronto, Ontario}
\email{isk@yorku.ca}
\thanks{}

\author{Michael C. Haslam}
\address{Department of Mathematics and Statistics}
\curraddr{York University, Toronto, Ontario}
\email{mchaslam@mathstat.yorku.ca}
\thanks{This research was partially supported by a grant given to MCH from the Natural Sciences and Engineering Research Council of Canada. This research has made use of NASA's Astrophysics Data System.}


\date{April 24, 2014}

\begin{abstract}
We use a dynamical systems approach based on the method of orthonormal frames to study the dynamics of a two-fluid, non-tilted Bianchi Type I cosmological model. Such a universe is anisotropic, spatially homogeneous, and spatially flat. In our model, one of the fluids is a fluid with bulk viscosity, while the other fluid assumes the role of a cosmological constant and represents nonnegative vacuum energy.  We begin by completing a detailed fixed-point analysis of the system which gives information about the local sinks, sources and saddles. We then proceed to analyze the global features of the dynamical system by using topological methods such as finding Lyapunov and Chetaev functions, and finding the $\alpha$- and $\omega$-limit sets using the LaSalle invariance principle. The fixed points found were a flat Friedmann-LeMa\^{\i}tre-Robertson-Walker (FLRW) universe with no vacuum energy, a de Sitter universe, a flat FLRW universe with both vacuum and non-vacuum energy, and a Kasner quarter-circle universe. We also show in this paper that the vacuum energy we observe in our present-day universe could actually be a result of the bulk viscosity of the ordinary matter in the universe, and proceed to calculate feasible values of the bulk viscous coefficient based on observations reported in the Planck data. We conclude the paper with some numerical experiments that shed further light on the global dynamics of the system.
\end{abstract}

\maketitle

\section{Introduction}
In this paper, we use a dynamical systems approach to investigate in detail the dynamics of a Bianchi Type I universe with a bulk viscous fluid and cosmological constant. Such a universe is spatially flat, spatially homogeneous, and anisotropic. Such a model may have considerable importance in present studies of cosmology given the recent results of the Planck measurements \cite{planckdata}, which suggest that the curvature of the spatial sections of the present-day universe is in agreement with spatial flatness. Moreover, Bianchi models are more general than the Friedmann-Lema\^{\i}tre-Robertson-Walker (FLRW) models and therefore can provide better descriptions of the early universe where viscous effects may have been dominant \cite{hervik}. One can then study the effects of viscosity on the dynamical evolution of the universe which we observe to be of FLRW-type today. As discussed by Coley and Wainwright \cite{coleywainwright}, cosmological models with single fluids are necessarily a simplification in the sense that they can only describe one epoch during the evolution of the universe. More general models can be constructed using two fluids with barotropic equations of state that are also comoving. One can then use these models to describe the transitions between different epochs in the universe's evolution, such as going from a radiation-dominated phase $(w = -1/3)$ to a matter-dominated phase $(w = 0)$. As discussed by Gr{\o}n and Hervik (Chapter 13, \cite{hervik}), viscous models have become of general interest in early-universe cosmologies largely in two contexts.  Firstly, in models where bulk viscous terms dominate over shear terms, the universe expands to a de Sitter-like state, which is a spatially flat universe neglecting ordinary matter, and including only a cosmological constant. Such models isotropize indirectly through the massive expansion. Secondly, in the absence of any significant heat flux, shear viscosity is found to play an important role in models of the universe at its early stages. In particular, neutrino viscosity is considered to be one of the most important factors in the isotropization of our universe. By also including a nonnegative cosmological constant in our model, and interpreting it to represent vacuum energy, we are also able to give a detailed description of the roles played by both viscosity and vacuum energy in the isotropization of our universe.

Bianchi cosmological models which contain a viscous fluid matter source in addition to a cosmological constant have been studied in detail several times. Lorenz-Petzold \cite{lorenzp} examined Bianchi Type I and V models in the the presence of perfect fluid matter with bulk viscosity and a nonzero cosmological constant.  Pradhan and Pandey \cite{pradhanpandey} studied Bianchi Type I magnetized cosmological models in the presence of a bulk viscous fluid in addition to a monotonically decreasing cosmological constant. Saha \cite{sahaI} studied the evolution of a Bianchi Type I universe with a viscous fluid and a cosmological constant. Pradhan, Srivastav, and Yadav \cite{pradhan2} studied Bianchi Type IX viscous models with a time-dependent positive cosmological constant. Belinch\'on \cite{belinchon} investigated the dynamics of a locally rotationally symmetric (LRS) Bianchi Type I universe with a bulk viscous fluid and a time-dependent cosmological constant. Pradhan, Jotania, and Rai \cite{pradhan3} studied Bianchi Type V cosmological models with bulk viscous fluid and a time-dependent cosmological constant. They also discussed some physical and geometrical aspects of such models. Pradhan and Pandey \cite{pradhan4} studied Bianchi Type I cosmological models with both shear and bulk viscosity and a monotonically decreasing cosmological constant. The authors considered the special case in which the expansion tensor only had two components. Saha and Rikhvitsky \cite{saha} analyzed a Bianchi Type I universe with a cosmological constant and dissipative processes due to viscosity. They showed that a positive cosmological constant leads to an ever-expanding universe. Singh and Kale \cite{singhkale2} studied Bianchi Type I, Kantowski-Sachs, and Bianchi Type III cosmological models containing as matter sources a bulk viscous fluid and non-constant gravitational and cosmological constants. Pradhan and Kumhar \cite{pradhankumhar} studied LRS Bianchi Type II models with bulk viscous fluid and a decaying cosmological constant. Mostafapoor and Gr\o n \cite{mostafgron} studied a Bianchi Type I universe with a cosmological constant and nonlinear viscous fluid. Sadeghi, Amani, and Tahmasbi \cite{sadeghi} investigated a Bianchi Type-VI cosmological model with a cosmological constant and viscous fluid. Barrow \cite{barrownuc2} showed that models of an inflationary universe driven by Witten strings in the very early universe are equivalent to the addition of bulk viscosity to perfect fluid cosmological models with zero curvature. In this work, Barrow considered the case where the bulk viscosity has a power-law dependence upon the matter density. It was shown that if the exponent is greater than $1/2$, there exist deflationary solutions which begin in a de Sitter state and evolve away from it asymptotically in the future. On the other hand, if this exponent is less than $1/2$, then solutions expand from an initial singularity towards a de Sitter state. Barrow  \cite{barrownuc3} also estimated the entropy production associated with anisotropy damping in the early universe by considering a Bianchi type I metric with an equilibrium radiation gas and anisotropic stresses produced by shear viscosity. It was shown that the shear viscosity based on kinetic theory has the general form of being proportional to the matter density and that the entropy production due to collisional transport is negligible in such a model.

All of the aforementioned papers use the metric approach (Page 39, \cite{ellis}) to obtain the dynamical evolution of the Bianchi model under consideration. The alternative approach which is based on the method of orthonormal frames pioneered by Ellis and MacCallum \cite{ellismac} in conjunction with dynamical systems theory is the path we take in this paper. Belinkskii and Khalatnikov \cite{belkhat} used phase-plane techniques to study a Bianchi Type I model under the influence of both shear and bulk viscosity.  Goliath and Ellis \cite{goliathellis} used dynamical systems methods to study FLRW, Bianchi Type I,  Bianchi Type II, and Kantoswki-Sachs models with a positive cosmological constant. Coley and van den Hoogen \cite{coleyvanV} analyzed in detail a Bianchi Type V model with viscosity, heat conduction, and a cosmological constant. They showed that all models that satisfy the weak energy condition isotropize. Coley, van den Hoogen, and Maartens \cite{coley} examined the full Israel-Stewart theory of bulk viscosity applied to dissipative FLRW models. Coley and Dunn \cite{coleydunn} used dynamical systems methods to study the evolution of a Bianchi Type V model with both shear and bulk viscosity. Burd and Coley \cite{burdcoley} examined using dynamical systems methods the effects of both bulk and shear viscosities upon the FLRW, Bianchi Type I, Bianchi Type V, and Kantowski-Sachs models. They found that these models were structurally stable under the introduction of bulk viscosity. Kohli and Haslam \cite{isk1} used dynamical systems methods to study the future asymptotic behavior of a Bianchi Type IV model containing both bulk and shear viscosity. Kohli and Haslam \cite{isk2} used dynamical systems methods to study a Bianchi Type I model containing bulk and shear viscosity in addition to a homogeneous magnetic field. 

With respect to dynamical systems methods in multi-fluid models, Stabell and Refsdal \cite{stabell} considered the dynamics of a two-fluid FLRW system consisting of dust and a cosmological constant. Phase plane methods were used by Madsen, Mimoso, Butcher, and Ellis \cite{madsenmimoso} to study the evolution of FLRW models in the presence of an arbitrary mixture of perfect fluids. Coley and Wainwright \cite{coleywainwright} examined orthogonal Bianchi and FLRW models in the presence of a two-fluid system. Ehlers and Rindler \cite{ehlersrind} studied in great detail three-fluid models, containing radiation, dust, and a cosmological constant. Recently, Barrow and Yamamoto\cite{barrowyama} considered a two-fluid system with one of the fluids representing a cosmological constant in their study of the instabilities of Bianchi Type IX Einstein static universes. For more details on the history of multi-fluid models, the interested reader should see Pages 53-55, 60-62, 171-172 and references therein of \cite{ellis}.

Despite all of the important aforementioned contributions, we feel it will be of considerable value to consider the dynamics of a Bianchi Type I universe with a viscous fluid and cosmological constant with respect to dynamical systems theory following the methods outlined in \cite{ellis} and \cite{barrowyama}. To the best of the authors' knowledge at the time of writing this paper, such an investigation has not been carried out in the literature. 

Throughout this paper, we assume a metric signature of $(-,+,+,+)$ and use geometrized units, where $8\pi G = c = 1$.

\section{The Evolution Equations}
We begin by describing the physical constituents of our two-fluid model. It can be shown \cite{isk1} that in the absence of heat conduction, the energy-momentum tensor of a fluid with both bulk and shear viscosity is given by
\begin{equation}
\label{Vab}
\mathcal{V}_{ab} = \left(\mu_{m} + p_{m}\right)u_{a} u_{b} + g_{ab} p_{m} - 3 \xi H h_{ab} - 2 \eta \sigma_{ab},
\end{equation}
where $\mu_{m}, p_{m}, \sigma_{ab}$, and $u_{a}$ represent the matter density, pressure, shear tensor, and fluid four-velocity respectively. Further, the quantities $\xi$ and $\eta$ denote the bulk and shear viscosity coefficients of the fluid matter source, $H$ denotes the Hubble parameter, and $h_{ab} \equiv u_{a} u_{b} + g_{ab}$ is the standard projection tensor corresponding to our assumed metric signature.

The second fluid in our model represents a cosmological constant, which can be modelled as a perfect fluid with barotropic equation of state $p_{\Lambda} = -\mu_{\Lambda}$. That is, the equation of state parameter is $w = -1$. The energy-momentum tensor for such a cosmological constant takes the simple form
\begin{equation}
\label{Lab}
\Lambda_{ab} = -g_{ab} \mu_{\Lambda},
\end{equation}
where $\mu_{\Lambda}$ in this case represents the vacuum energy density corresponding to the cosmological constant.

Assuming the matter in our model described by Eq. \eqref{Vab} assumes a barotropic equation of state $p_{m} = w \mu_{m}$, where in general, $-1 \leq w \leq 1$, using the definitions
\begin{equation}
\mu_{m} = \mathcal{V}_{ab} u^{a} u^{b}, \quad p_{m} = \frac{1}{3} h^{ab} \mathcal{V}_{ab}, \quad \pi_{ab} = h^{c}_{a} h^{d}_{b} \mathcal{V}_{cd} - p h_{ab},
\end{equation}
we find that
\begin{equation}
\label{mattereqs}
p_{m} = w \mu_{m} - 3 \xi H, \quad \pi_{ab} = -2 \eta \sigma_{ab},
\end{equation}
where $\pi_{ab}$ represents the total anisotropic stress of the fluid. 

To write down the Einstein field equations as a dynamical system, it is necessary that we express the above variables in their expansion-normalized form \cite{ellis}, thus introduce the definitions
\begin{equation}
\label{normdefs}
\Omega_{m} = \frac{\mu_{m}}{3H^2}, \quad \Omega_{\Lambda} = \frac{\mu_{\Lambda}}{3H^2}, \quad P_{m} = \frac{p_{m}}{3H^2}, \quad P_{\Lambda} = \frac{p_{\Lambda}}{3H^2}, \quad \Pi_{ab} = \frac{\pi_{ab}}{H^2}.
\end{equation}
Following \cite{coley}, \cite{belkhat}, and \cite{barrownuc2}, we define the expansion-normalized form of the bulk and shear viscosity coefficients as
\begin{equation}
\label{viscdefs}
\frac{\xi}{3H} = \xi_{0} \Omega_{m}^{a}, \quad \frac{\eta}{3H} = \eta_{0} \Omega_{m}^{b},
\end{equation}
where $\xi_{0}$ and $\eta_{0}$ denote the bulk and shear viscosity parameters and are taken to be nonnegative. In addition, the exponents $a$ and $b$ are also assumed to be nonnegative. We will discuss the problem of choosing values for these exponents in the next section when deriving equilibrium points of the dynamical system.

In deriving the evolution equations, we essentially follow \cite{ellis} and note that we consider Bianchi models relative to a group-invariant orthonormal frame $\left\{\mathbf{n}, \mathbf{e}_{k}\right\}$, $(k = 1,2,3)$ where $\mathbf{n}$ is the the unit normal to the group orbits. Since $\mathbf{n}$ is tangent to a hypersurface-orthogonal congruence of geodesics, these equations are obtained by assuming that all variables are only functions of time, the motion of the matter is along geodesics, and there is no vorticity. The basic dynamical variables are then
\begin{equation}
\label{basicvars}
\left(H, \sigma_{ab}, n_{ab}, a_{a}\right),
\end{equation}
where $n_{ab}$ and $a_{a}$ classify and represent the spatial curvature of the specific Bianchi model under question.

If we now apply the definitions in Eq. \eqref{viscdefs} to the basic variables in Eq. \eqref{basicvars}, we obtain the expansion-normalized evolution equations as given in \cite{hewittbridsonwainwright} and \cite{elliscosmo} as:
\begin{eqnarray}
\label{eq:evolutionsys1}
\Sigma_{ij}' &=& -(2-q)\Sigma_{ij} + 2\epsilon^{km}_{(i}\Sigma_{j)k}R_{m} - \mathcal{S}_{ij} + \Pi_{ij} \nonumber \\
N_{ij}' &=& qN_{ij} + 2\Sigma_{(i}^{k}N_{j)k} + 2 \epsilon^{km}_{(i}N_{j)k}R_{m} \nonumber \\
A_{i}' &=& qA_{i} - \Sigma^{j}_{i}A_{j} + \epsilon_{i}^{km}A_{k} R_{m}\nonumber \\
\Omega' &=& (2q - 1)\Omega - 3P - \frac{1}{3}\Sigma^{j}_{i}\Pi^{i}_{j} + \frac{2}{3}A_{i}Q^{i} \nonumber \\
Q_{i}' &=& 2(q-1)Q_{i} - \Sigma_{i}^{j}Q_{j} - \epsilon_{i}^{km}R_{k}Q_{m} + 3A^{j}\Pi_{ij} + \epsilon_{i}^{km}N_{k}^{j}\Pi_{jm}.
\end{eqnarray}
These equations are subject to the constraints
\begin{eqnarray}
\label{eq:constraints1}
N_{i}^{j}A_{j} &=& 0 \nonumber \\
\Omega &=& 1 - \Sigma^2 - K \nonumber \\
Q_{i} &=& 3\Sigma_{i}^{k} A_{k} - \epsilon_{i}^{km}\Sigma^{j}_{k}N_{jm}.
\end{eqnarray}
In the expansion-normalized approach, $\Sigma_{ab}$ denotes the kinematic shear tensor, and describes the anisotropy in the Hubble flow, $A_{i}$ and $N^{ij}$ describe the spatial curvature, while $\Omega^{i}$ and $R^{i}$ describe the relative orientation of the shear and spatial curvature eigenframes and energy flux respectively. Further the \emph{prime} denotes differentiation with respect to a dimensionless time variable $\tau$ such that
\begin{equation}
\frac{dt}{d\tau} = \frac{1}{H}.
\end{equation}

Considering a Bianchi Type I model, by definition, we have that
\begin{equation}
N_{ij} = \mbox{diag}\left(0,0,0\right), \quad A_{i} = R_{i} = Q_{i} = 0.
\end{equation}

In this paper, we only consider the case where the fluid matter source has nonzero bulk viscosity, and therefore set $\eta_{0} = 0$. The importance of this assumption has been discussed in for example, \cite{barrownuc2}. Therefore, upon considering Eqs. \eqref{mattereqs}, \eqref{normdefs}, \eqref{viscdefs}, \eqref{eq:evolutionsys1}, and \eqref{eq:constraints1}, we obtain the evolution equations for our system as
\begin{eqnarray}
\label{evo1}
\Sigma_{+}' &=& \Sigma_{+} \left(q - 2\right), \\
\Sigma_{-}' &=&  \Sigma_{-} \left(q - 2\right), \\
\Omega_{m}' &=& \Omega_{m} \left(2q - 1 -3w\right) + 9 \xi_{0} \Omega_{m}^{a}, \\
\label{evo2}
\Omega_{\Lambda}' &=& 2\left(q+1\right) \Omega_{\Lambda},
\end{eqnarray}
where $q$ is the deceleration parameter which can be obtained by setting the Raychaudhuri equation (Eq (1.90) in \cite{ellis}) to the general evolution equation for $H$ (Eq. (5.8) in \cite{ellis}) and then solving for $q$. Proceeding in this manner gives
\begin{equation}
q = 2\left(\Sigma_{+}^2 + \Sigma_{-}^2\right) + \frac{1}{2}\left[\Omega_{m}\left(1+3w\right)\right] - \frac{9}{2}\xi_{0}\Omega_{m}^{a} - \Omega_{\Lambda}.
\end{equation}
The equations \eqref{evo1}-\eqref{evo2} are subject to the constraint
\begin{equation}
\label{friedmann}
\Omega_{m} + \Omega_{\Lambda} + \Sigma_{+}^2 + \Sigma_{-}^2 = 1,
\end{equation}
which is just the generalized Friedmann equation (Eq. (1.92) in \cite{ellis}) in expansion-normalized form.  Also, note that in equations \eqref{evo1}-\eqref{evo2} we have made use of the notation
\begin{equation}
\Sigma_{+} = \frac{1}{2}\left(\Sigma_{22} + \Sigma_{33}\right), \quad \Sigma_{-} = \frac{1}{2\sqrt{3}}\left(\Sigma_{22} - \Sigma_{33}\right),
\end{equation}
such that $\Sigma^2 \equiv \Sigma_{+}^2 + \Sigma_{-}^2$.
We note that although it is possible to reduce the dynamical system dimension through the use of the constraint Eq. \eqref{friedmann}, in general, as stated in \cite{goliathellis}, for Bianchi models, the resulting reduced system will be a projection of the full state space, and the complete information expressed by the dynamical system may be lost. We therefore follow the strategy in \cite{barrowyama} and do not reduce the state space in our analysis of the dynamical system.

\section{Stability Analysis of the Dynamical System}
With the evolution and constraint equations in hand, we will now perform a detailed analysis of the equilibrium points of the dynamical system. The system of equations \eqref{evo1}-\eqref{evo2} is a nonlinear, autonomous system of ordinary differential equations, and can be written as
\begin{equation}
\label{eq:basedef}
\mathbf{x}' = \mathbf{f(x)},
\end{equation}
where $\mathbf{x} = \left[\Sigma_{+}, \Sigma_{-}, \Omega_{m}, \Omega_{\Lambda}\right] \in \mathbf{R}^{4}$, and  the vector field $\mathbf{f(x)}$ denotes the right-hand-side of the dynamical system. The dynamical system also exhibits some symmetries, specifically ones that leave the system invariant with respect to spatial inversions of the dynamical variables. These are given by
\begin{eqnarray}
\label{restr1}
\phi_{1}&:& \left[\Sigma_{+}, \Sigma_{-}, \Omega_{m}, \Omega_{\Lambda}\right] \rightarrow  \left[-\Sigma_{+}, \Sigma_{-}, \Omega_{m}, \Omega_{\Lambda}\right], \\
\label{restr2}
\phi_{2}&:& \left[\Sigma_{+}, \Sigma_{-}, \Omega_{m}, \Omega_{\Lambda}\right] \rightarrow  \left[\Sigma_{+}, -\Sigma_{-}, \Omega_{m}, \Omega_{\Lambda}\right].
\end{eqnarray}
These symmetries imply that we can take
\begin{equation}
\label{sigmaconstr}
\Sigma_{\pm} \geq 0.
\end{equation}
In addition, based on the physical constraints of having nonnegative energy density, we make the assumption that
\begin{equation}
\label{sigmaconstr2}
\Omega_{m} \geq 0, \quad \Omega_{\Lambda} \geq 0.
\end{equation}
Tavakol (Chapter 4, \cite{ellis}) discusses a simple way to obtain the invariant sets of a dynamical system. Let us consider a dynamical system $\dot{x} = v(x), \quad x \in \mathbb{R}^{4}$. Let $Z: \mathbb{R}^{4} \to \mathbb{R}$ be a $C^{1}$ function such that $Z' = \alpha Z$, where $\alpha: \mathbb{R}^{4}$ is a continuous function. Then the subsets of $\mathbb{R}^{4}$ defined by $Z > 0$, $Z = 0$, and $Z < 0$ are invariant sets of the flow of the dynamical system. Applying this proposition to our dynamical system in combination with the symmetries found above, we see that $\Sigma_{\pm} \geq 0$ and $\Omega_{\Lambda} \geq 0$ constitute invariant sets of the dynamical system.

Following \cite{abrahammarsden}, we first note that the vector field $\mathbf{f(x)}$ is clearly at least $C^{1}$ on $M = \mathbb{R}^{4}$. We call a point $\mathbf{m_{0}}$ an equilibrium point of $\mathbf{f(x)}$ if $\mathbf{f(m_{0})} = 0$. Let $(U, \phi)$ be a chart on $M$ with $\phi(\mathbf{m_{0}}) = \mathbf{x}_{0} \in \mathbb{R}^{4}$, and let $\mathbf{x} = \left(\Sigma_{+}, \Sigma_{-}, \Omega_{m}, \Omega_{\Lambda}\right)$ denote coordinates in $\mathbb{R}^{4}$.  Then, the linearization of  $\mathbf{f(x)}$ at $\mathbf{m_{0}}$ in these coordinates is given by
\begin{equation}
\label{lin1}
\left(\frac{\partial \mathbf{f(x)}^{i}}{\partial x^{j}}\right)_{\mathbf{x} = \mathbf{x}_{0}}
\end{equation}
It is a remarkable fact of dynamical systems theory that if the point $\mathbf{m_{0}}$ is hyperbolic, then there exists a neighborhood $N$ of $\mathbf{m_{0}}$ on which the flow of the system $F_{t}$ is topologically equivalent to the flow of the linearization Eq. \eqref{lin1}. This is the theorem of Hartman and Grobman \cite{ellis}. That is, in $N$, the orbits of the dynamical system can be deformed continuously into the orbits of Eq. \eqref{lin1}, and the orbits are therefore topologically equivalent. We use the following convention when discussing the stability properties of the dynamical system. If all eigenvalues $\lambda_{i}$ of Eq. \eqref{lin1} satisfy $Re(\lambda_{i}) < 0 (Re(\lambda_{i}) > 0)$, $\mathbf{m_{0}}$ is local sink (source) of the system. If the point $\mathbf{m_{0}}$ is neither a local source or sink, we will call it a saddle point.

Solving for the equilibrium points, we first obtain three types of flat FLRW-type solutions:
\begin{eqnarray}
\label{F1}
\mathcal{F}_{1} &:&  \Sigma_{+} = 0, \quad \Sigma_{-} = 0, \quad \Omega_{m} = 1, \quad \Omega_{\Lambda} = 0, \\
\label{deSitter}
\mathcal{D} &:& \Sigma_{+} = 0, \quad \Sigma_{-} = 0, \quad \Omega_{m} = 0, \quad \Omega_{\Lambda} = 1, \\
\label{mixFLRW}
\mathcal{F}_{2} &:&  \Sigma_{+} = 0, \quad \Sigma_{-} = 0, \quad \Omega_{m} = \left(\frac{1+w}{3\xi_{0}} \right)^{\frac{1}{a-1}}, \quad \Omega_{\Lambda} = 1 - \Omega_{m},
\end{eqnarray}
where $\mathcal{D}$ is the de Sitter solution. For $\mathcal{F}_{2}$, based on the physical constraints of having nonnegative bulk viscosity and energy densities, we must have additionally that
\begin{equation}
\label{F2constr}
-1 < w \leq 1, \quad 0 < \xi_{0} \leq \frac{1+w}{3}.
\end{equation}
It is also interesting to see that for $\mathcal{F}_{2}$,
\begin{equation}
\lim_{a \to \pm \infty} \Omega_{m} = 1 \Rightarrow \Omega_{\Lambda} = 0.
\end{equation}
Therefore, it can be said that the point $\mathcal{F}_{1}$ exists in the extreme limit with respect to the exponent $a$ of the point $\mathcal{F}_{2}$.

In the special case where we additionally have that $\xi_{0} = 0$, we obtain a Kasner circle equilibrium point:
\begin{eqnarray}
\label{K}
\mathcal{K} &:&  \Sigma_{+}^2 + \Sigma_{-}^2 = 1, \quad \Omega_{m} = 0, \quad \Omega_{\Lambda} = 0.
\end{eqnarray}
We should note that this is a special type of Kasner circle, in that, it is actually a Kasner quarter-circle. This is evident due to the restrictions described in Eq. \eqref{sigmaconstr}.

Analyzing the stability of $\mathcal{F}_{1}$, we note that the eigenvalues of Eq. \eqref{lin1} at this point are found to be
\begin{equation}
\label{eigsF1}
\lambda_{1} = \lambda_{2} =  \frac{3}{2} (-1+w-3 \xi_{0} ), \quad \lambda_{3} = 1+3 w-9 \xi_{0}, \quad \lambda_{4} = 3 (1+w-3 \xi_{0} ).
\end{equation}
Therefore, the point $\mathcal{F}_{1}$ is a local sink if
\begin{equation}
\left(-1 \leq w \leq 1\right) \wedge \left( \xi_{0} > \frac{1+w}{3}\right),
\end{equation}
and is a saddle point if
\begin{equation}
\left(-1<w<-\frac{1}{3}\wedge 0\leq \xi_{0} <\frac{1+w}{3}\right)\vee \left(-\frac{1}{3}\leq w\leq 1\wedge \frac{1}{9} (1+3 w)<\xi_{0} <\frac{1+w}{3}\right)
\end{equation}
or
\begin{equation}
\left(-\frac{1}{3}<w<1\wedge0\leq \xi_{0} <\frac{1}{9} (1+3 w)\right)\vee \left(w=1\wedge0<\xi_{0} <\frac{4}{9}\right).
\end{equation}
As can be shown the point $\mathcal{F}_{1}$ is never a source of the system.

Analyzing the stability of $\mathcal{F}_{2}$, we note that the eigenvalues of Eq. \eqref{lin1} at this point are found to be
\begin{eqnarray}
\label{eigsF2}
\lambda_{1} &=& -2+3 (1+w) z^{\frac{1}{-1+a}}-9 \left(z^{\frac{1}{-1+a}}\right)^a \xi_{0}, \nonumber \\
\lambda_{2} &=&  \frac{3}{2} \left[-2+(1+w) z^{\frac{1}{-1+a}}-3 \left(z^{\frac{1}{-1+a}}\right)^a \xi_{0} \right], \nonumber \\
\lambda_{3} &=& \lambda_{2}, \nonumber \\
\label{eigsF2end}
\lambda_{4} &=& z^{\frac{1}{1-a}} \left[-3 (1+w) z^{\frac{1}{-1+a}}+6 (1+w) z^{\frac{2}{-1+a}}-9 \left(z^{\frac{1}{-1+a}}\right)^a \left(z^{\frac{1}{-1+a}}+a \left(-1+z^{\frac{1}{-1+a}}\right)\right) \xi_{0} \right], \nonumber \\
\end{eqnarray}
where we have defined
\begin{equation}
z \equiv \frac{1+w}{3 \xi_{0}}.
\end{equation}
Clearly, finding equilibrium points of the system by solving $\mathbf{f(m_{0})} = 0$ is a difficult task for general $a$ in Eqs. \eqref{evo1}-\eqref{evo2}. We must therefore choose values for $a$ beforehand, and then perform the fixed-point analysis. Our choices for these values are not completely arbitrary. Belinksii and Khalatnikov \cite{belkhat} and Barrow \cite{barrownuc2} have both presented arguments for physically relevant choices for these exponents. In particular, following \cite{belkhat} and \cite{barrownuc2}, we note that it is reasonable to consider $a \leq 1/2$ for the early universe, where we expect viscous effects to play a significant role in its dynamical evolution.

\subsection{The case: $a=0$}
Setting $a=0$ in Eq. \eqref{mixFLRW}, we obtain:
\begin{eqnarray}
\mathcal{F}_{3}&:& \Sigma_{+} = 0, \quad \Sigma_{-} = 0, \quad \Omega_{m} = \frac{3\xi_{0}}{1+w}, \quad \Omega_{\Lambda} = \frac{1 + w - 3\xi_{0}}{1+w}.
\end{eqnarray}

Analyzing the stability of $\mathcal{F}_{3}$,  the eigenvalues in Eqs. \eqref{eigsF2} - \eqref{eigsF2end} are found to be
\begin{equation}
\label{eigsF3}
\lambda_{1} = -2, \quad \lambda_{2} = -3\left(1 + w - 3\xi_{0}\right), \quad \lambda_{3} = \lambda_{4} = -3.
\end{equation}
Therefore, $\mathcal{F}_{3}$ is a local sink if
\begin{equation}
\label{F3cond1}
\left(-1 < w \leq 1\right) \wedge \left(0 \leq \xi_{0} < \frac{1+w}{3}\right),
\end{equation}
and is a saddle point if
\begin{equation}
\left(-1 < w \leq 1\right) \wedge \left(0 \leq \xi_{0} > \frac{1+w}{3}\right).
\end{equation}
Clearly, the region corresponding to $\mathcal{F}_{3}$ being a saddle point is unphysical since it violates Eq. \eqref{F2constr}. Therefore, $\mathcal{F}_{3}$ can never be a saddle point. Additionally, $\mathcal{F}_{3}$ can never be a source of the system.

\subsection{The case: $a=1/2$}
Setting $a=1/2$ in Eq. \eqref{mixFLRW}, we obtain:
\begin{eqnarray}
\label{eigsF4}
\mathcal{F}_{4}&:& \Sigma_{+} = 0, \quad \Sigma_{-} = 0, \quad \Omega_{m} = \frac{9 \xi_{0}^2}{\left(1+w\right)^2}, \quad \Omega_{\Lambda} = \frac{1+2 w+w^2-9 \xi_{0} ^2}{(1+w)^2}.
\end{eqnarray}

The eigenvalues in Eqs. \eqref{eigsF2} - \eqref{eigsF2end} are found to be
\begin{equation}
\lambda_{1} = \lambda_{2} = -3, \quad \lambda_{3} = -2, \quad \lambda_{4} = -\frac{3 \left(1+2 w+w^2-9 \xi_{0} ^2\right)}{2 (1+w)}.
\end{equation}
It is easy to show that this point is always a local sink if
\begin{equation}
\label{F3cond2}
-1 < w \leq 1, \quad 0 \leq \xi_{0} < \frac{1+w}{3}.
\end{equation}
It is in fact true that even if $\xi_{0} = (1+w)/3$ such that $\lambda_{4} = 0$, $\mathcal{F}_{4}$ will still be a local sink, since it will be a normally hyperbolic point (Chapter 4, \cite{ellis}). 


We note that Eq. \eqref{lin1} is not defined at $\mathcal{D}$ or $\mathcal{K}$, so linearization techniques will not help us in determining the stability of these points. However, in the next section, we describe some other techniques that will help us determine the global stability of these points.

\section{Global Results}
Complementing the preceding fix-point analysis, we wish to obtain some information about the asymptotic behavior of the dynamical system as $\tau \to \pm \infty$. To accomplish this, we make use of the LaSalle Invariance Principle, and the methods of finding Lyapunov and Chetaev functions.  According to Theorem 4.11 in \cite{ellis}, the LaSalle Invariance Principle for $\omega$-limit sets is stated as follows. Consider a dynamical system $\mathbf{x}' = \mathbf{f(x)}$ on $\mathbb{R}^{n}$, with flow $\phi_{t}$. Let $S$ be a closed, bounded and positively invariant set of $\phi_{t}$ and let $Z$ be a $C^{1}$ monotone function. Then $\forall$ $\mathbf{x}_{0} \in S$, we have that $\omega(\mathbf{x}_{0}) \subseteq \left\{\mathbf{x} \in S | Z' = 0\right\}$, where $Z' = \nabla Z \cdot \mathbf{f}$. The extended LaSalle Invariance Principle for $\alpha$-limit sets can be found in Proposition B.3. in \cite{hewwain}. To use this principle, one simply considers $S$ to be a closed, bounded, and negatively invariant set. Then $\forall$ $\mathbf{x}_{0} \in S$, we have that $\alpha(\mathbf{x}_{0}) \subseteq \left\{\mathbf{x} \in S | Z' = 0\right\}$, where $Z' = \nabla Z \cdot \mathbf{f}$. 

Following Pages 24 and 25 of \cite{arnolddiff}, we note that a differentiable function $Z$ is called a \emph{Lyapunov function} for a singular point $\mathbf{x}_{0}$ of a vector field $\mathbf{(f(x)}$ if $Z$ is defined on a neighborhood of $\mathbf{x_{0}}$ and has a local minimum at this point, and the derivative of $Z$ along $\mathbf{f(x)}$ is nonpositive. Then a singular point of a differentiable vector field for which a Lyapunov function exists is stable. Further, a differentiable function $Z$ is called a \emph{Chetaev function} for a singular point $\mathbf{x_{0}}$ of a vector field $\mathbf{f(x)}$ if $Z$ is defined on a domain $W$ whose boundary contains $\mathbf{x}_{0}$, the part of the boundary of $W$ is strictly contained in a sufficiently small ball with its center $\mathbf{x}_{0}$ removed is a piecewise-smooth, $C^{1}$ hypersurface along which $\mathbf{f(x)}$ points into the interior of the domain, that is,
\begin{equation}
Z(\mathbf{x}) \to 0, \mbox{ as } \mathbf{x} \to \mathbf{x}_{0}, \quad \mathbf{x} \in W; \quad Z > 0, \quad \nabla Z \cdot \mathbf{f(x)} > 0 \in W.
\end{equation}
A singular point of a $C^{1}$ vector field for which a Chetaev function exists is unstable.
%

Let us first consider the invariant set
\begin{equation}
\label{inv1}
S_{1} = \left\{\Sigma_{+} = \Sigma_{-} = \Omega_{\Lambda} = 0\right\}.
\end{equation}
We will also define the function
\begin{equation}
Z_{1} = \Omega_{m},
\end{equation}
such that within $S_{1}$ we have
\begin{equation}
Z_{1}' = \left(-1 + \Omega_{m}\right) \left[\Omega_{m} \left(1 + 3w\right) - 9 \xi_{0} \Omega_{m}^{a}\right].
\end{equation}
This function is monotone if for example $\Omega_{m} = 1$, and strictly monotone decreasing if
\begin{equation}
\Omega_{m} < 1, \quad w > 3\xi_{0} - \frac{1}{3}.
\end{equation}
By the LaSalle invariance principle, for any orbit $\Gamma \in S_{1}$, we have that
\begin{equation}
\omega(\Gamma) \subseteq \left\{\Omega_{m} = 1\right\},
\end{equation}
which corresponds precisely to the FLRW equilibrium point as given in Eq. \eqref{F1}. Therefore, the global future asymptotic state of the system corresponding to the invariant set Eq. \eqref{inv1} is a flat FLRW universe with $\Omega_{m} = 1$.                                                                                                                                                                                                                                                                                                                                                                                                                                                                 

Consider now the function
\begin{equation}
Z_{2} = \Omega_{\Lambda}^2 + 1.
\end{equation}
Let us define the neighborhood of the de Sitter equilibrium point $\mathcal{D}$ as the open ball
\begin{equation}
D_{r}(\mathcal{D}) = \left[\Sigma_{+}^2 + \Sigma_{-}^2 + \Omega_{m}^2 + \left(\Omega_{\Lambda} - 1\right)^2 \right]^{1/2} < r,
\end{equation}
where $r > 0$ is the radius of this ball in $\mathbb{R}^{4}$. That this is an open neighborhood of $\mathcal{D}$ is proven in Section 2.2 in \cite{marsdencalc}. Clearly the function $Z_{2}$ is defined on $D_{r}(\mathcal{D})$. One can also show that the point $\Omega_{\Lambda} = 1$ is a local minimum of $Z_{2}$ in $D_{r}(\mathcal{D})$. Furthermore, from computing
\begin{equation}
Z_{2}' = -4\Omega_{\Lambda}^2\left(-1 + \Omega_{\Lambda}\right),
\end{equation}
one can see that on $D_{r}(\mathcal{D})$, $Z_{2}' \leq 0$ as long as $\Omega_{\Lambda} \leq 1$. Therefore, $Z_{2}$ is a Lyapunov function corresponding to $\mathcal{D}$, and as a result, $\mathcal{D}$ is stable.

Consider now the invariant set
\begin{equation}
\label{inv3}
S_{2} = \left\{\Sigma_{+} = 0, \quad \Sigma_{-} = 0 \quad 0 < \Omega_{m} < 1, \quad 0 < \Omega_{\Lambda} < 1\right\}.
\end{equation}
We now define the function
\begin{dmath}
\label{Z3func}
Z_{3} = -9 \left[\frac{1}{3} (1+w)^2 \Omega_{\Lambda}^3-\frac{1}{2} (1+w)^2 \Omega_{\Lambda}^4+\frac{1}{5} (1+w)^2 \Omega_{\Lambda}^5+\frac{6 (1+w) \xi_{0}  (1-\Omega_{\Lambda})^{2+a} \left(2+2 (2+a) \Omega_{\Lambda}+\left(6+5 a+a^2\right) \Omega_{\Lambda}^2\right)}{(2+a) (3+a) (4+a)}-\frac{9 \xi_{0} ^2 (1-\Omega_{\Lambda})^{1+2 a} \left(1+\Omega_{\Lambda}+2 a \Omega_{\Lambda}+\left(1+3 a+2 a^2\right) \Omega_{\Lambda}^2\right)}{(1+a) (1+2 a) (3+2 a)}\right].
\end{dmath}
Noting that
\begin{equation}
\frac{d Z_{3}}{d \tau} = \frac{d Z_{3}}{d \Omega_{\Lambda}} \frac{d \Omega_{\Lambda}}{d\tau},
\end{equation}
we have that within the invariant set $S_{2}$, 
\begin{equation}
\label{Z3deriv}
Z_{3}' =18 (-1+\Omega_{\Lambda}) \Omega_{\Lambda}^3 \left[-1+3 \xi_{0}  (1-\Omega_{\Lambda})^a+w (-1+\Omega_{\Lambda})+\Omega_{\Lambda}\right]^2,
\end{equation}
which is strictly monotone decreasing along orbits in $S_{2}$, specifically, for $0 < \Omega_{\Lambda} < 1$. Note that based on Eq. \eqref{Z3func}, we must have that $a \neq -2, -3, -4, -1, -1/2, -3/2$. Also, in deriving Eq. \eqref{Z3deriv}, we used Eq. \eqref{friedmann} to write $\Omega_{m} = 1 - \Omega_{\Lambda}$. The point $\mathcal{F}_{2}$ as given in Eq. \eqref{mixFLRW} is contained in the set $S_{2}$. Furthermore, we have that
\begin{equation}
Z_{3}'(\mathcal{F}_{2}) = 0, \quad \mathcal{F}_{2} \in S_{2}, \quad a > 1 \in \mathbb{Z}.
\end{equation}
Therefore, by the LaSalle invariance principle, we must have that for all orbits $\Gamma \in S_{2}$,
\begin{equation}
\omega(\Gamma) \subseteq \left\{0 < \Omega_{\Lambda} < 1 \right\},
\end{equation}
which is another possible future asymptotic state of the dynamical state and corresponds to the equilibrium point $\mathcal{F}_{2}$.

Let us now consider the domain
\begin{equation}
W = \left\{\Sigma_{+}^2 + \Sigma_{-}^2 <1, \quad 0< \Omega_{m} < 1, \quad 0< \Omega_{\Lambda} < 1\right\}.
\end{equation}
The boundary of this domain is given by
\begin{equation}
\bar{W} \backslash W = \left\{\Sigma_{+}^2 + \Sigma_{-}^2 = 1\right\} \cup \left\{\Omega_{m} = 0\right\} \cup \left\{\Omega_{\Lambda} = 0\right\} \cup \left\{\Omega_{m} = 1\right\} \cup \left\{\Omega_{\Lambda} = 1 \right\}.
\end{equation}
Clearly, the Kasner circle equilibrium point $\mathcal{K}$ is contained in $\bar{W} \backslash W$. Let us now define a function on $W$,
\begin{equation}
Z_{4}(\mathbf{x}) = \frac{\Omega_{m}}{1 - \Omega_{\Lambda}}.
\end{equation}
We note some important properties of this function. First,
\begin{equation}
\lim_{\mathbf{x} \to \mathcal{K}} Z_{4} (\mathbf{x}) = 0, \quad \mathbf{x} \in W.
\end{equation}
Further, we have that
\begin{equation}
Z_{4}(\mathbf{x}) > 0, \quad \mathbf{x} \in W.
\end{equation}
Using Eqs. \eqref{evo1}-\eqref{evo2} and  computing $Z_{4}'(\mathbf{x})$, we obtain
\begin{equation}
Z_{4}'(\mathbf{x}) = \frac{\Omega_{m} \left[-1 + 4 \Sigma_{-}^2 + 4 \Sigma_{+}^2 + \Omega_{\Lambda} + \Omega_{m} + 3w\left(-1 + \Omega_{\Lambda} + \Omega_{m}\right)\right]}{\left(\Omega_{\Lambda} -1\right)^2}.
\end{equation}
Considering the special case of $w = -1/3$, this equation takes the form
\begin{equation}
Z_{4}'(\mathbf{x}) = \frac{4 \Omega_{m} \left(\Sigma_{+}^2 + \Sigma_{-}^2\right)}{\left(\Omega_{\Lambda} -1\right)^2},
\end{equation}
which is strictly positive everywhere in $W$. Therefore, we have just proven that for the special case $w = -1/3$, $Z_{4}$ is a Chetaev function and corresponds to $\mathcal{K}$. This implies that the Kasner equilibrium point corresponding to $w = -1/3$ is unstable. We were only able to prove instability for this case. However, we conjecture based on extensive numerical experiments (see Fig. \ref{fig6}) that $\mathcal{K}$ is unstable for all values $-1 \leq w \leq 1$ and $\xi_{0} = 0$. 

\section{Bifurcations and Orbits}
\begin{figure}[H]
\begin{center}
\label{fig1}
\includegraphics[scale = 0.50]{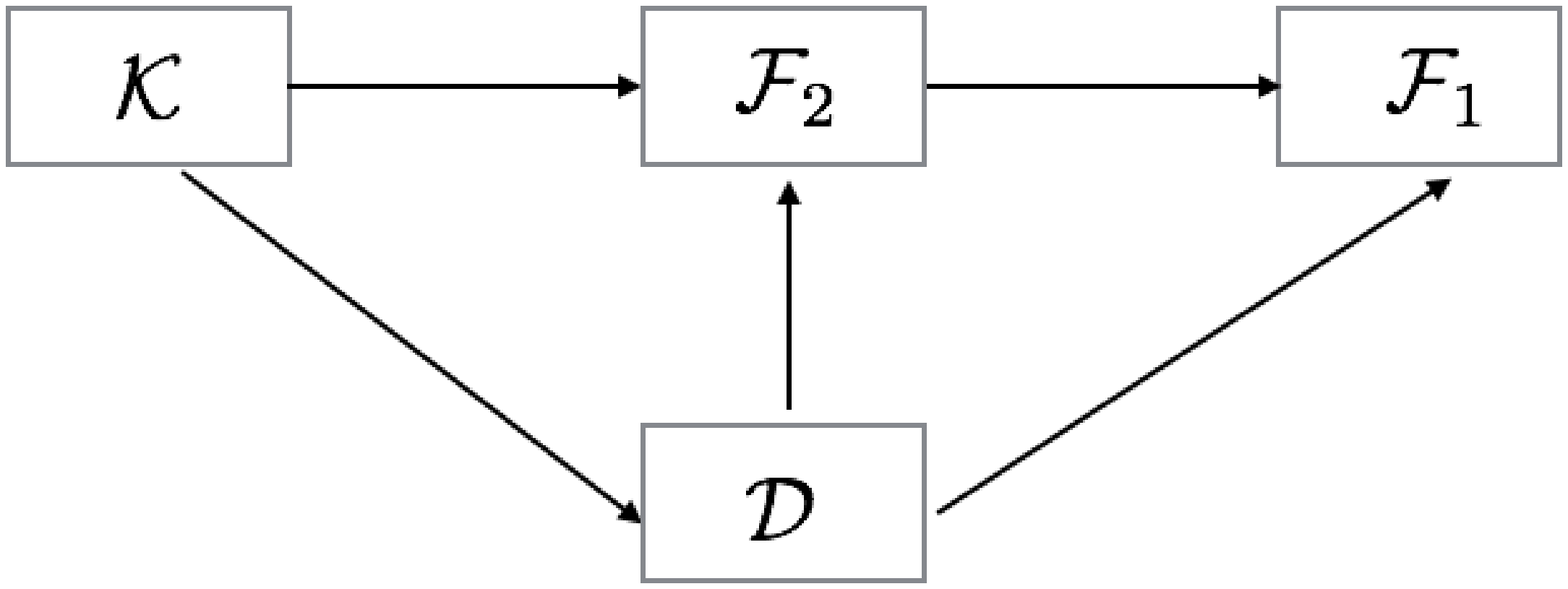}
\end{center}
\caption{The possible bifurcations that occur in the dynamical system as a result of changing the bulk viscosity coefficient $\xi_{0}$, the equation of state parameter $w$ or both.} 
\end{figure}

With the stability analysis completed in the previous section, we now will describe bifurcations that occur in the dynamical system. These occur by changing values of either the bulk viscosity coefficient $\xi_{0}$, the equation of state parameter $w$ or both. These bifurcations are displayed in Fig. \ref{fig1}. 

There also exists a finite heteroclinic sequence (Page 104, \cite{ellis}) when $\xi_{0} = 0$ and $-1 < w < -1/3$, and is given by
\begin{equation}
\label{het1}
\mathcal{K} \rightarrow \mathcal{F}_{1} \rightarrow \mathcal{F}_{2}.
\end{equation}

When $w = -1$ and $\xi_{0} = 0$, there exists a heteroclinic orbit:
\begin{equation}
\mathcal{K} \rightarrow \mathcal{D}.
\end{equation}
When $\xi_{0} = 0$ and $-1 < w \leq 1$, we have the following heteroclinic orbit:
\begin{equation}
\label{het2}
\mathcal{K} \rightarrow \mathcal{F}_{2}.
\end{equation}
There are some interesting things to note about the bifurcations, heteroclinic sequence and orbits described in this section. The Kasner equilibrium point, $\mathcal{K}$ is a state according to Eq. \eqref{K} that has no matter/energy or vacuum energy whatsoever, that is, $\Omega_{m} = \Omega_{\Lambda} = 0$. Yet, from this empty state, we see that in the case of the bifurcations, an increase in the bulk viscosity causes the universe to evolve towards a FLRW universe either with no vacuum energy or with a mix of matter and vacuum energy, which represents our universe today. A question that may be related to this phenomenon is what mechanism generates matter or particle creation not only in generic Bianchi type I spacetimes, but near a Kasner-type state specifically. This problem was first studied by Zeldovich \cite{zeldovich}, where he showed that spontaneous particle production could occur near a Kasner singularity. In related work, Parker \cite{parker} gave conditions for particle creation near an isotropic Friedmann-type singularity. Berger \cite{berger2} studied quantum graviton creation in a general spatially homogeneous, anisotropic three-torus solution of Einstein's equations. Ahmed and Mondal \cite{ahmedmondal} studied the Hawking radiation of Dirac particles in a Kasner spacetime. They showed that the anisotropy gives rise to particle creation. Further, Harko and Mak \cite{harkomak2} also investigated in some detail the effects of matter creation on the evolution and dynamics of a Bianchi type I model. Related to this, Barrow \cite{barrownuc3} studied the entropy production associated with anisotropy damping in the era of grand unification in the early universe. The general consensus in all of these investigations is that particle creation leads to the isotropization of an anisotropic spacetime. 

Further, according to the bifurcation sequences and finite heteroclinic sequences given in Eqs. \eqref{het1} and \eqref{het2}, we see that it is completely possible to go from an empty state to an FLRW state bypassing the de Sitter state altogether, which can have some interesting implications with regards to cosmic inflation. These results show that it is a valid to ask the question whether the universe can go from an initial empty state to the universe we have today without undergoing the standard inflationary epoch. That is, it may be that what we consider to be inflation with respect to an inflaton field may actually be an expansionary epoch driven by bulk viscosity. Barrow \cite{barrownuc2} investigated this matter in quite some detail. He found that when the bulk viscous exponent is larger than $1/2$ it is no longer guaranteed that an asymptotic de Sitter state will occur. In addition, when the exponent is greater than or equal to $1$, the asymptotic de Sitter state is replaced with a Weyl curvature singularity. Belinskii and Khalatnikov \cite{belkhat} also concluded in their study of the Bianchi type I universe with viscosity, that viscous effects alone showed an ``essential isotropizing action''. The relationship between the presence of bulk viscosity and the existence of an asymptotically stable de Sitter universe was also investigated in \cite{chimento}, \cite{zimdahl1}, \cite{zimdahl2}, and \cite{barrow4}.

\section{Connections with Observations}
The Planck team \cite{planckdata} recently calculated based on observations that 
\begin{equation}
\label{planckresults}
\Omega_{\Lambda} = 0.6825, \quad \Omega_{m} = 0.3175.
\end{equation}
This configuration is precisely modelled by our $\mathcal{F}_{2}$ equilibrium point given in Eq. \eqref{mixFLRW}, which we found to be a local sink:
\begin{equation*}
\Sigma_{+} = 0, \quad \Sigma_{-} = 0, \quad \Omega_{m} = \left(\frac{1+w}{3\xi_{0}} \right)^{\frac{1}{a-1}}, \quad \Omega_{\Lambda} = 1 - \Omega_{m}.
\end{equation*}
Given these facts and the fact that it is conjectured that the very early universe consisted of incoherent radiation with matter equation of state $w = 1/3$, or stiff matter with equation of state $w = 1$ (Page 99, \cite{elliscosmo}), we can ask the question what values of $\xi_{0}$ in the early universe could have led to the state described by Eqs. \eqref{planckresults} and \eqref{mixFLRW} today. To answer this question, we follow \cite{belkhat} and note that for the early universe, $a \leq 1/2$.  The interesting thing is that $\mathcal{F}_{2}$ is a future asymptotic state of the dynamical system as the eigenvalue computations show in Eqs. \eqref{eigsF3} and \eqref{eigsF4}. Therefore, choices of $w$ and $\xi_{0}$ in the early universe within the acceptable range of values as described in Eq. \eqref{F3cond1} would yield a future state that is similar to what we observe today. 

As a consequence of the aforementioned arguments, we now present some solutions of the system of equations
\begin{eqnarray}
\label{gensystem}
\left(\frac{1+w}{3\xi_{0}} \right)^{\frac{1}{a-1}} = 0.3175, \quad 1 -\left(\frac{1+w}{3\xi_{0}} \right)^{\frac{1}{a-1}} = 0.6825,
\end{eqnarray}
where $-1 < w \leq 1$, $\quad 0 < \xi_{0} \leq (1+w)/3$ as in Eq. \eqref{F2constr}.

For $a=0$, we obtain the parametrized solution
\begin{equation}
\label{a0solution}
0 < \xi_{0} \leq \frac{127}{600}, \quad w = \frac{1}{127} \left(-127 + 1200 \xi_{0}\right).
\end{equation}
It is important to recall that in our two-fluid model, the equation of state parameter $w$ describes the non-vacuum energy. It is conjectured that the majority of the non-vacuum energy/matter in the early universe consisted of radiation with equation of state parameter $w = 1/3$ (Page 98, \cite{elliscosmo}). Solving for $\xi_{0}$ in Eq. \eqref{a0solution} with $w = 1/3$, yields
\begin{equation}
\label{result1}
\xi_{0} = \frac{127}{900} \approx 0.1411, \quad w = 1/3, \quad a = 0.
\end{equation}
Solving for $\xi_{0}$ in Eq. \eqref{a0solution} with $w=1$ yields
\begin{equation}
\label{result11}
\xi_{0} = \frac{127}{600} \approx 0.2117, \quad w = 1, \quad a = 0.
\end{equation}

In this case where $a = 1/2$, we obtain the parameterized solution
\begin{equation}
\label{a1/2solution}
0<\xi_{0} \leq \frac{\sqrt{127}}{30}, \quad w= -1+\frac{60 \xi _{0}}{\sqrt{127}}.
\end{equation}
Solving for $\xi_{0}$ in Eq. \eqref{a1/2solution} with $w = 1/3$, yields
\begin{equation}
\label{result2}
\xi_{0} = \frac{\sqrt{127}}{45} \approx 0.2504, \quad w = 1/3, \quad a = \frac{1}{2}.
\end{equation}
Solving for $\xi_{0}$ in Eq. \eqref{a1/2solution} with $w=1$ yields
\begin{equation}
\label{result22}
\xi_{0} = \frac{\sqrt{127}}{30} \approx 0.3756, \quad w = 1, \quad a = \frac{1}{2}.
\end{equation}
Therefore, our model shows that an early-universe configuration with bulk viscosity values as computed in Eqs. \eqref{result1}, \eqref{result11}, \eqref{result2}, \eqref{result22} could lead to the mixture of vacuum and non-vacuum energy that we observe today. Further, our calculation of the $\mathcal{F}_{2}$ equilibrium point further shows that it is quite possible that the vacuum energy that exists in our universe today could be the result of some bulk viscous effect of the ordinary matter in the early universe or at the present time. The connections between bulk viscosity and vacuum energy have been explored in \cite{fenglishen}, \cite{singhsinghbali}, \cite{arbababdel} \cite{2005GReGr..37.2039B} \cite{2006PhLB..633....1R}, \cite{2006GReGr..38..495F},  \cite{2007PhRvD..76j3516C}, \cite{2009PhRvD..79j3521L}, \cite{2003PhRvD..67h3520D}, \cite{2007PhRvD..75d3521W},\cite{2011JCAP...09..026G}, \cite{2013PhRvD..88l3504V}, \cite{2012IJTP...51.2771D}, and \cite{2008MPLA...23.1372M}, though the calculations presented above to the best of the authors' knowledge are new and have not been reported before in the literature.

\section{Numerical Solutions}
To complement both the fixed-point and abstract topological analysis in the previous section, we now present some numerical solutions to the dynamical system Eqs. \eqref{evo1}-\eqref{evo2}. Initial conditions were chosen to satisfy the constraint equations \eqref{friedmann}, \eqref{sigmaconstr}, and\eqref{sigmaconstr2}, and are represented in the numerical experiments by asterisks. Furthermore, we note that the numerical solutions were completed over sufficiently long time intervals $\left(0 \leq \tau \leq 1000\right)$, but in some cases we present the solutions over shorter time intervals for clarity. 

\begin{figure}[H]
\begin{center}
\caption{This figure shows the dynamical system behavior for $\xi_{0} = 127/900$, $w = 1/3$, and $a = 0$. The circle denotes the equilibrium point $\mathcal{F}_{3}$. This precise case corresponds to Eq. \eqref{result1}.} 
\label{fig2}
\includegraphics[scale = 0.40]{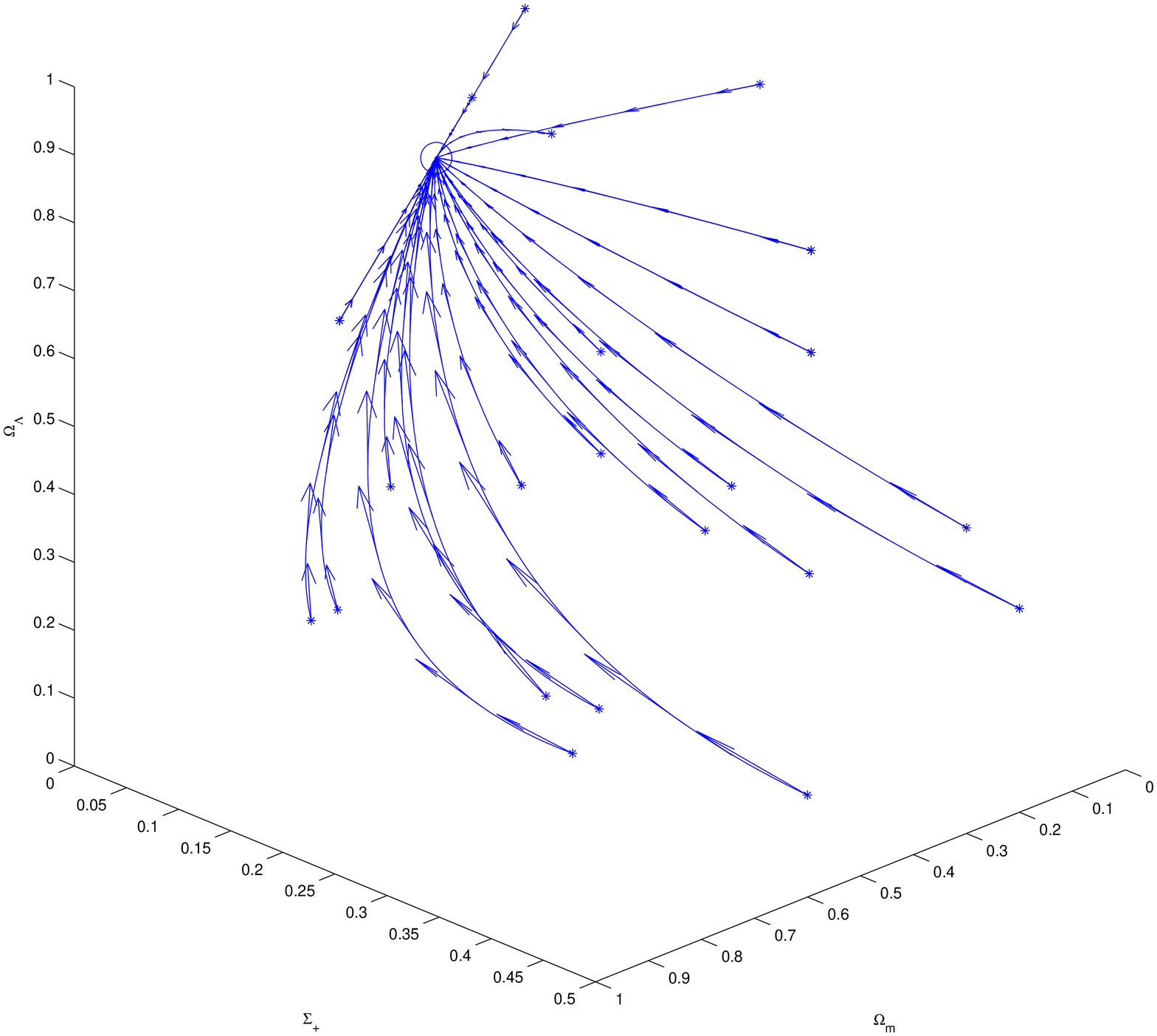}
\end{center}
\end{figure}

\begin{figure}[H]
\begin{center}
\caption{This figure shows the dynamical system behavior for $\xi_{0} = 127/600$, $w = 1$, and $a = 0$. The circle denotes the equilibrium point $\mathcal{F}_{3}$. This precise case corresponds to Eq. \eqref{result11}.} 
\label{fig3}
\includegraphics[scale = 0.40]{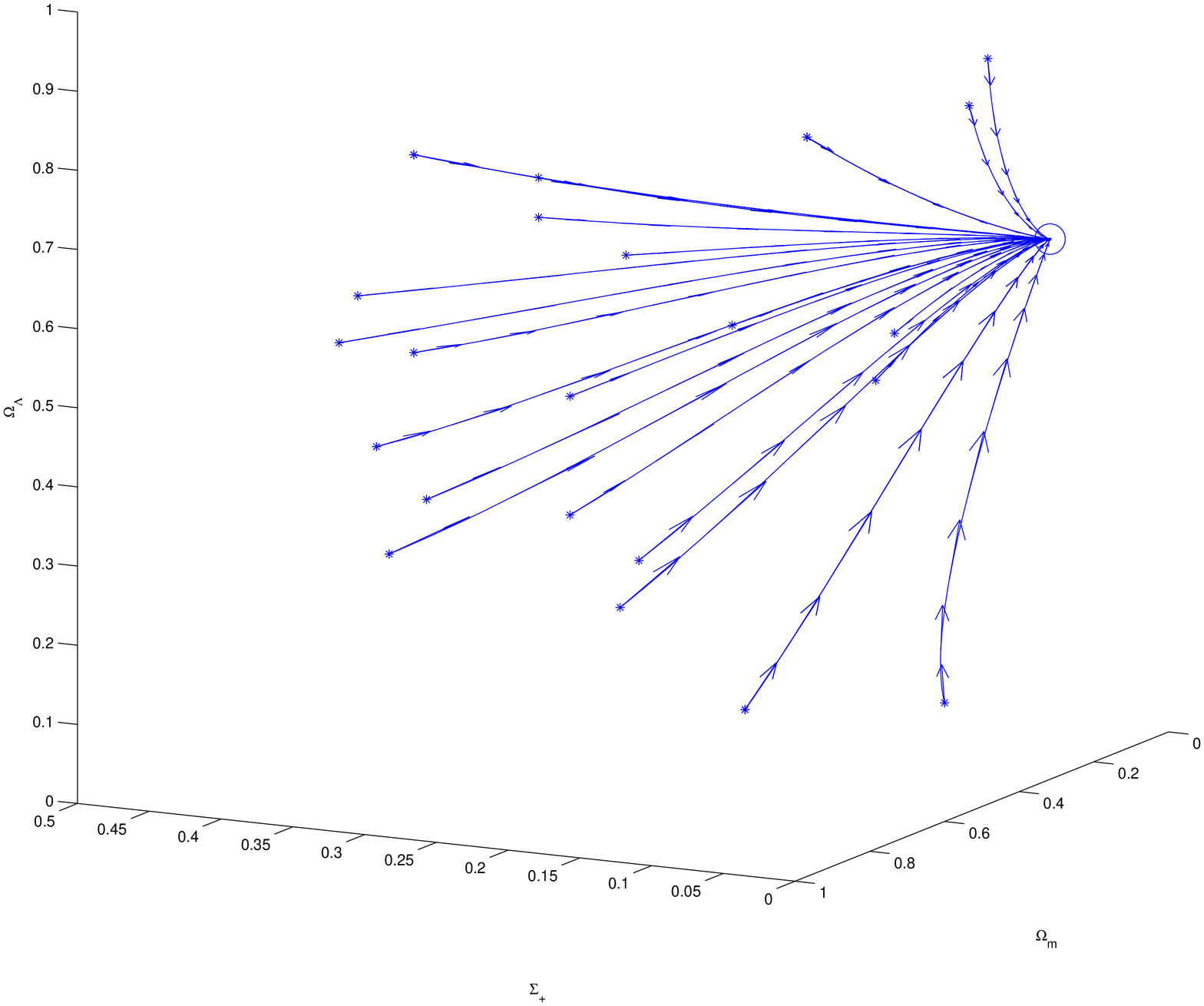}
\end{center}
\end{figure}

We display in Figs. \ref{fig2} and \ref{fig3} the results of numerical experiments that show that $\mathcal{F}_{3}$ is indeed a local sink of the system.

\begin{figure}[H]
\begin{center}
\caption{This figure shows the dynamical system behavior for $\xi_{0} = \sqrt{127}/45$, $w = 1/3$, and $a = 1/2$. The circle denotes the equilibrium point $\mathcal{F}_{4}$. This precise case corresponds to Eq. \eqref{result2}.} 
\label{fig4}
\includegraphics[scale = 0.40]{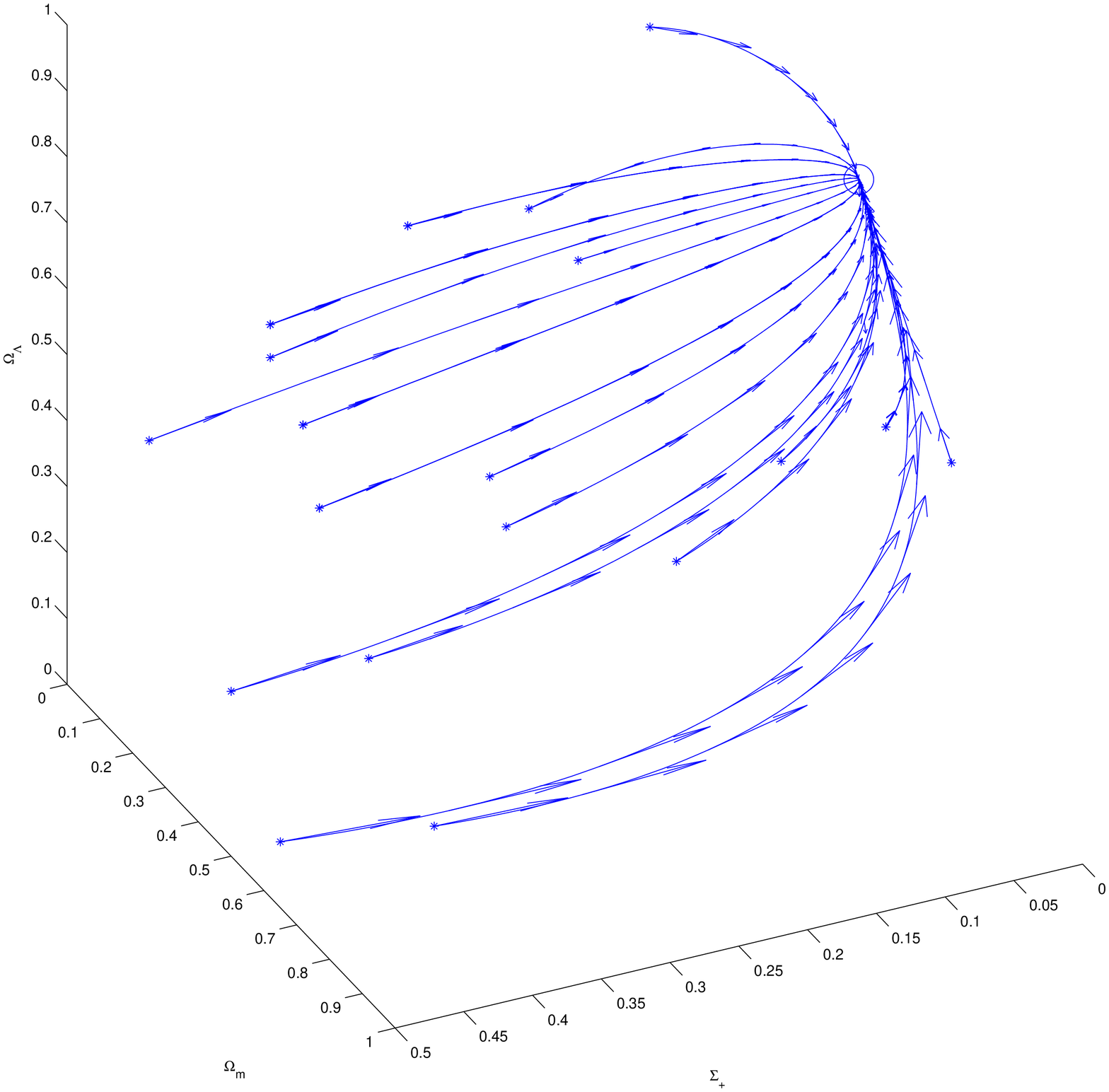}
\end{center}
\end{figure}

\begin{figure}[H]
\begin{center}
\caption{This figure shows the dynamical system behavior for $\xi_{0} = \sqrt{127}/30$, $w = 1$, and $a = 1/2$. The circle denotes the equilibrium point $\mathcal{F}_{4}$. This precise case corresponds to Eq. \eqref{result22}.} 
\label{fig5}
\includegraphics[scale = 0.40]{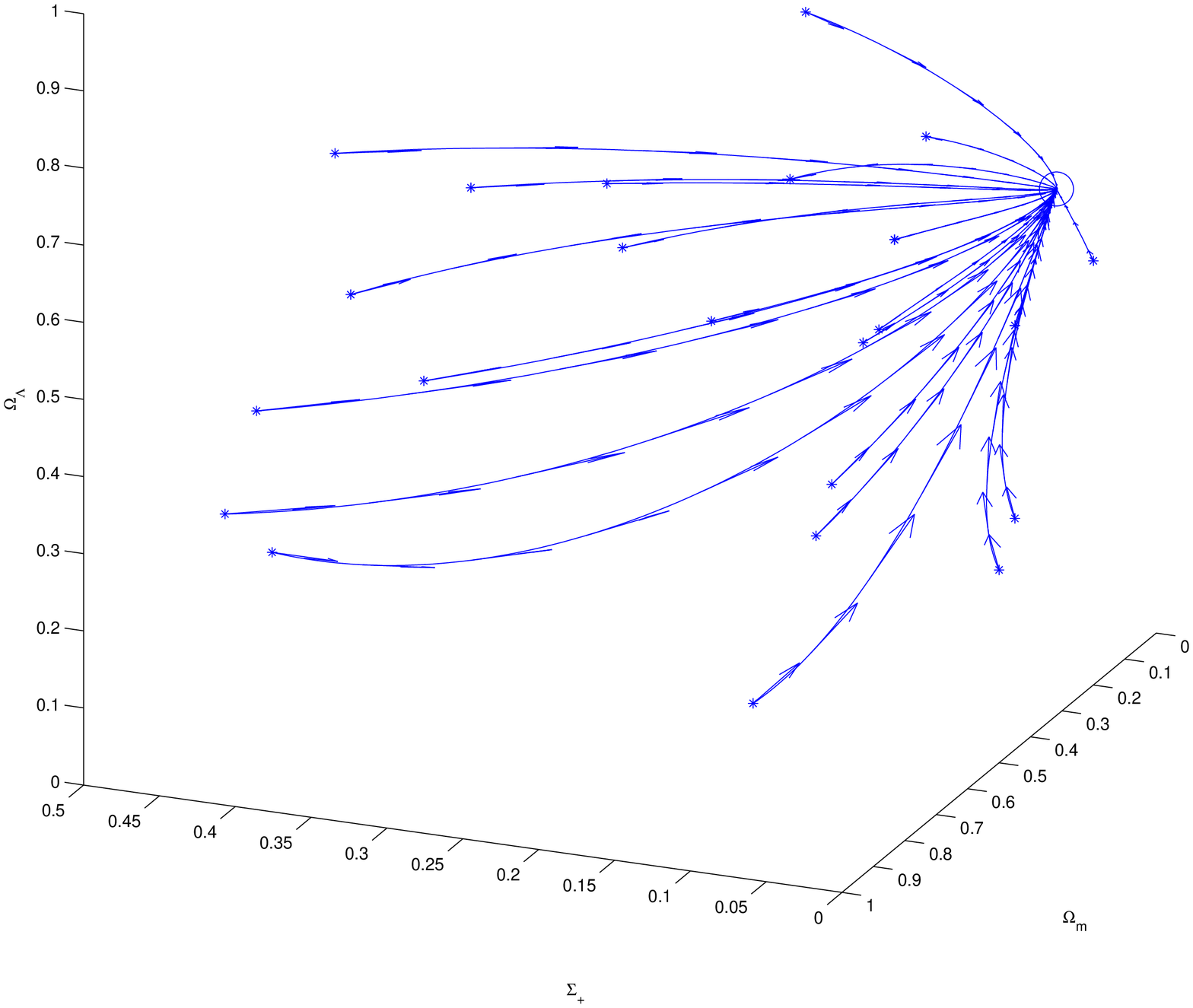}
\end{center}
\end{figure}

We display in Figs. \ref{fig4} and \ref{fig5} the results of numerical experiments that show that $\mathcal{F}_{4}$ is indeed a local sink of the system.

\begin{figure}[H]
\begin{center}
\caption{This figure shows the dynamical system behavior for $\xi_{0} = 0$, and $w = -1, -1/3, 0, 1/3, 1$. One can see that in each case, $\mathcal{K}$ is a source of the dynamical system. The case $w = -1/3$ clearly corresponds to our analysis at the end of Section IV. Note that the boundary of circles corresponds to the Kasner quarter-circle.} 
\label{fig6}
\includegraphics[scale = 0.60]{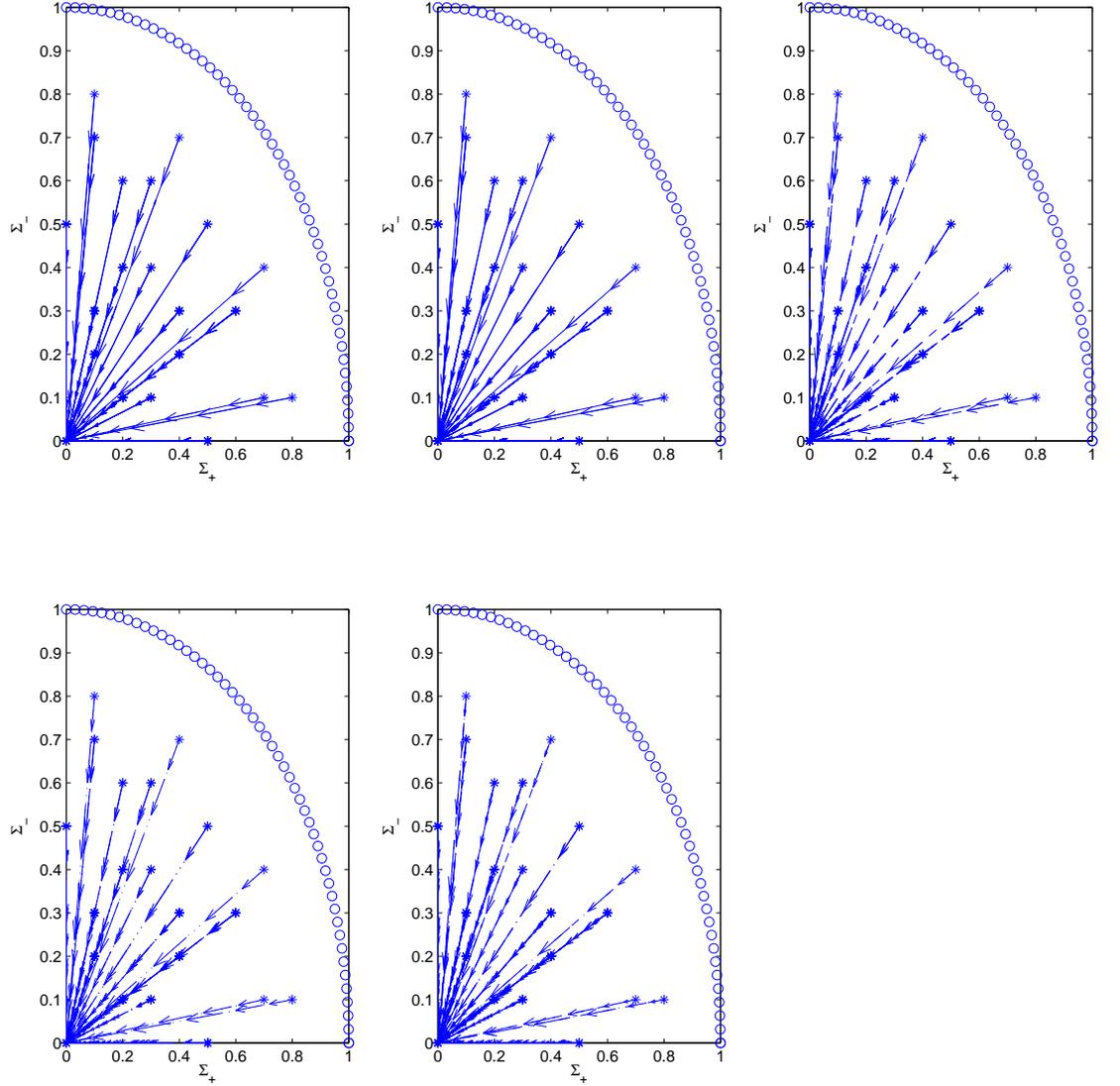}
\end{center}
\end{figure}

We display in Fig. \ref{fig6} the results of several numerical experiments that show $\mathcal{K}$ as a source of the dynamical system.

\begin{figure}[H]
\begin{center}
\caption{This figure shows the dynamical system behavior for $\xi_{0} = 1/2$ and $w = 1/3$, which denotes radiation. The circle denotes the equilibrium point $\mathcal{F}_{1}$. Clearly this point is a local sink of the dynamical system.} 
\label{fig7}
\includegraphics[scale = 0.40]{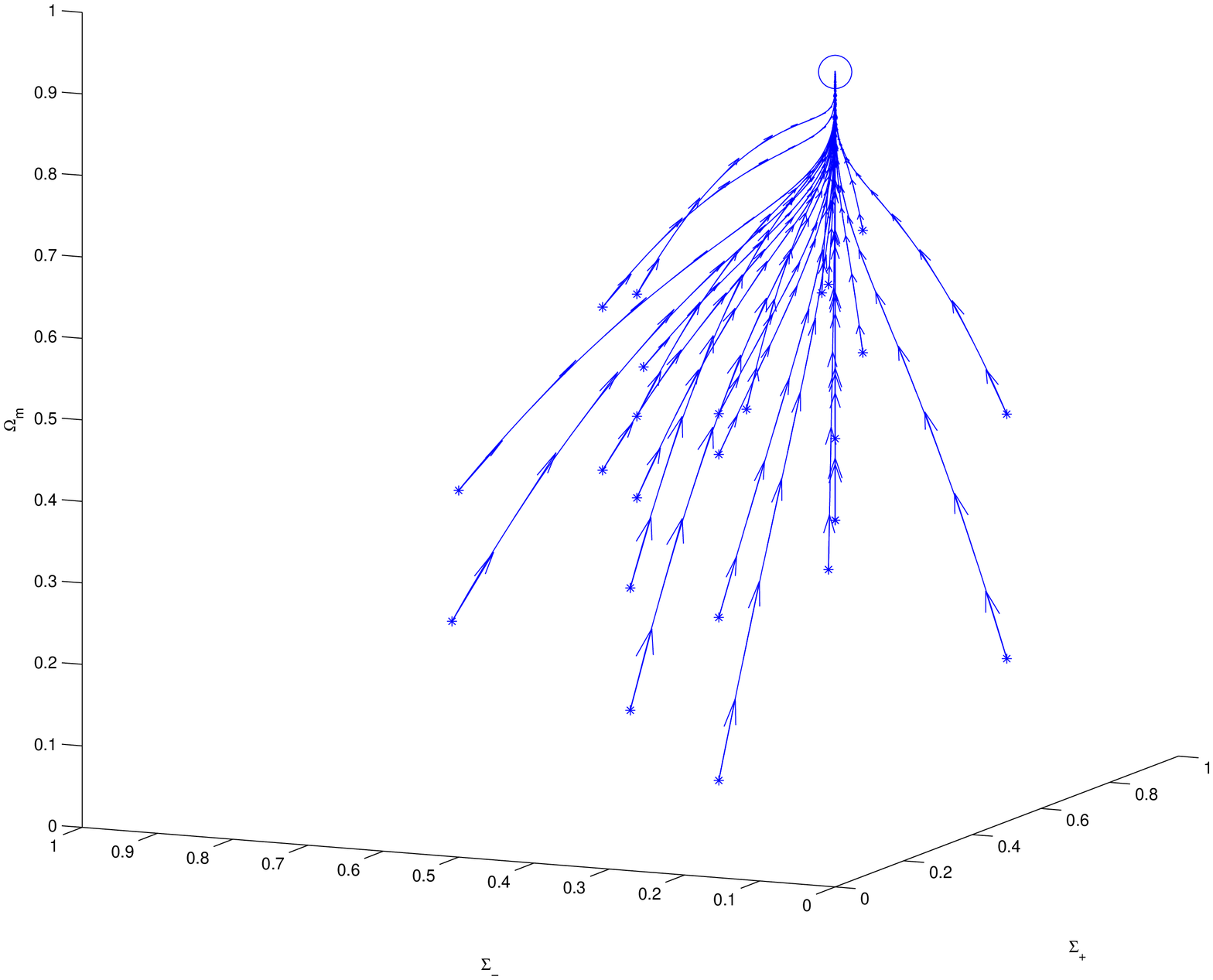}
\end{center}
\end{figure}

\begin{figure}[H]
\begin{center}
\caption{This figure shows the dynamical system behavior for $\xi_{0} = 0.34$ and $w = 0$, which denotes dust. The circle denotes the equilibrium point $\mathcal{F}_{1}$. Clearly this point is a local sink of the dynamical system.} 
\label{fig8}
\includegraphics[scale = 0.40]{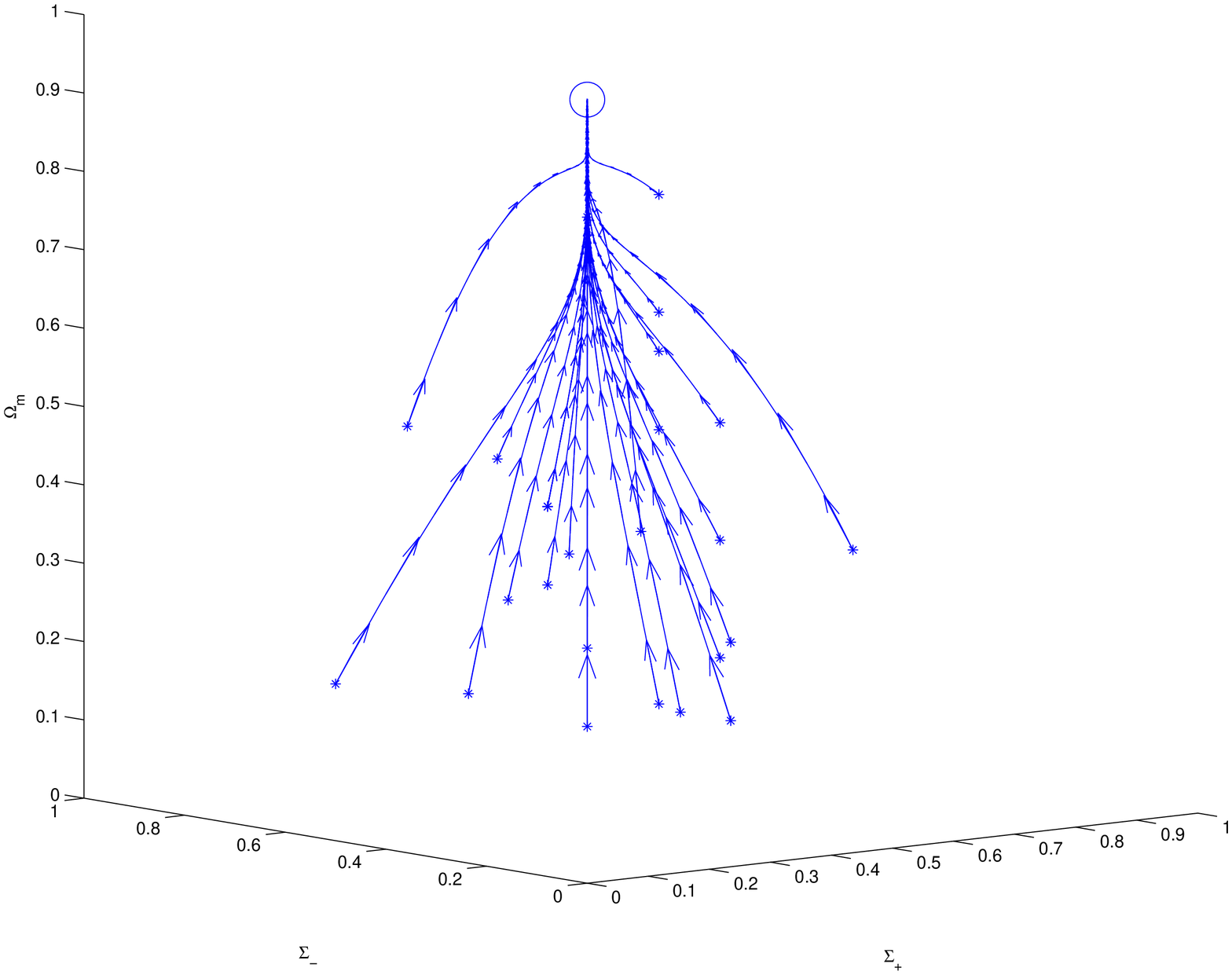}
\end{center}
\end{figure}

\begin{figure}[H]
\begin{center}
\caption{This figure shows the dynamical system behavior for $\xi_{0} = 0.45$ and $w = 0.325$, which denotes a dust-radiation mixture. The circle denotes the equilibrium point $\mathcal{F}_{1}$. Clearly this point is a local sink of the dynamical system.} 
\label{fig9}
\includegraphics[scale = 0.40]{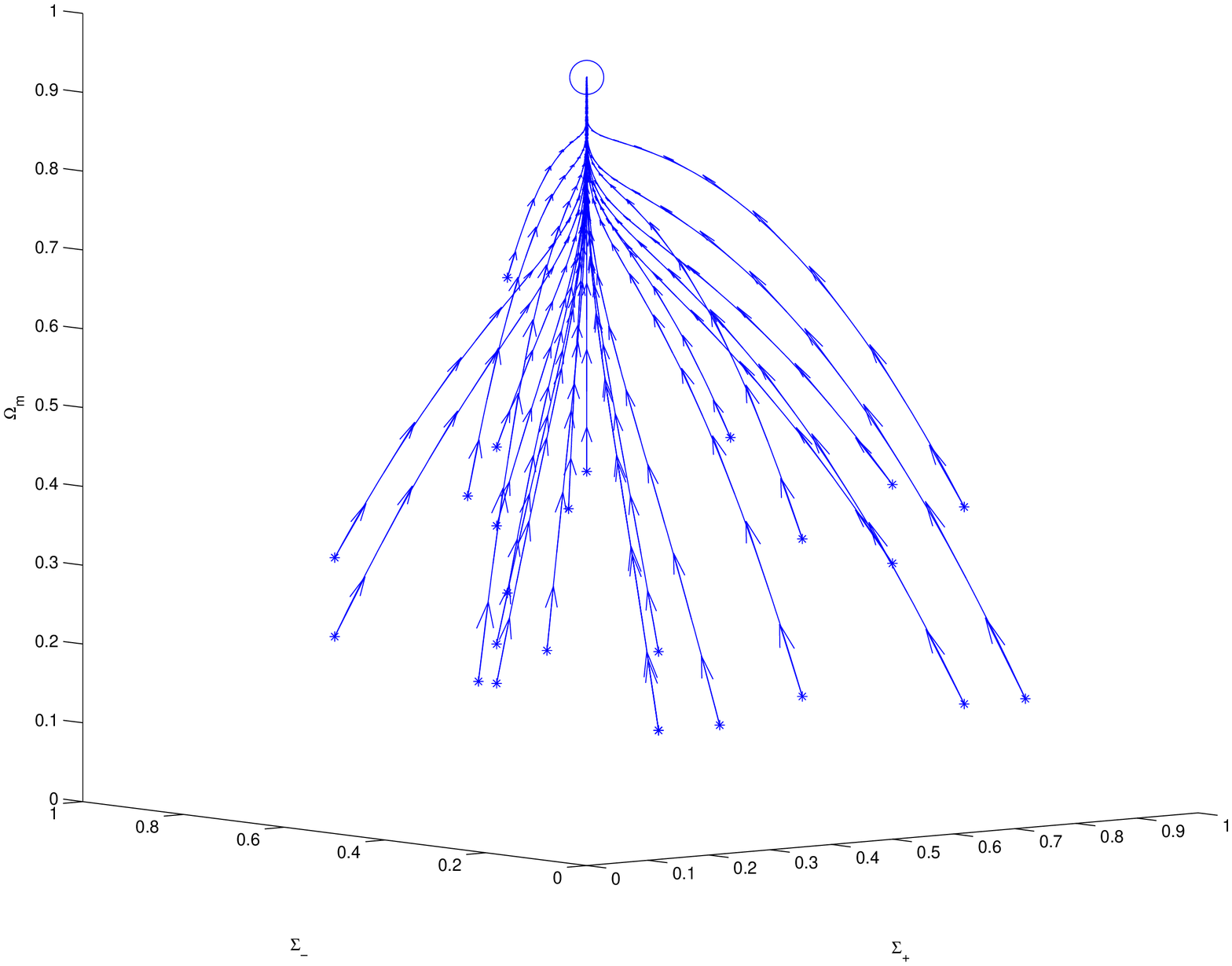}
\end{center}
\end{figure}

We display in Figs. \ref{fig7}, \ref{fig8}, \ref{fig9} the results of several numerical experiments that show $\mathcal{F}_{1}$ as a local sink of the dynamical system.

\section{Conclusions}
We have presented in this paper a comprehensive analysis of the dynamical behavior of a Bianchi Type I two-fluid model with bulk viscosity and a cosmological constant. We began by completing a detailed fixed-point analysis of the system which gave information about the local sinks, sources and saddles. We then proceeded to analyze the global features of the dynamical system by using topological methods such as finding Lyapunov and Chetaev functions, and finding the $\alpha$- and $\omega$-limit sets using the LaSalle invariance principle. 

The fixed points found were a flat FLRW universe with no vacuum energy and only energy due to ordinary matter, a de Sitter universe, a mixed FLRW universe with both vacuum and non-vacuum energy, and a Kasner quarter-circle universe. We found conditions for which the former three were local sinks of the system, that is, future asymptotic states, and where the latter was a source of the system, that is, a past asymptotic state. 

The flat FLRW universe solution we found with both vacuum and non-vacuum energy is clearly of primary importance with respect to modelling the present-day universe, especially in light of the recently-released Planck data. In fact, using this Planck data we gave possible conditions for which a non-zero bulk viscosity in the early universe could have lead to some of the conditions described in the Planck data in the present epoch. In particular, since we found that this equilibrium point is a local sink of the dynamical system, all orbits approach this equilibrium point in the future. Therefore, there exists a time period for which our cosmological model will isotropize and be compatible with present-day observations of a high degree of isotropy of the cosmic microwave background in addition to the existence of both vacuum and non-vacuum energy.

\bibliographystyle{amsplain}
\bibliography{sources}

\end{document}